\documentclass{article}
\usepackage{latexsym}
\usepackage{amsmath}
\usepackage{amssymb}
\title{Efficient Simulation of Heavy Quark Vacuum Polarization}
\author{Matthew Nobes\footnote{
    Institute for High Energy Phenomenology, Cornell University, 
      Ithaca, New York, USA, 14853-5001.  Email: nobes@lepp.cornell.edu}}
\begin{document}
  
  \maketitle

  \begin{abstract}
    We outline a simple way to include heavy quark vacuum polarization in 
    lattice QCD simulations.  The method, based on effective field theory, 
    requires only a trivial modification of the gluon
    action and has no impact on simulation times.  We assess the 
    range of validity for this procedure, and the impact that it may have.
  \end{abstract}

  \section{Introduction}
  
  The inclusion of light quark vacuum polarization in lattice QCD Monte Carlo 
  simulations has long been a major problem.  With the development of 
  improved staggered fermions \cite{lepage98} it has become possible to do 
  high-precision lattice QCD, with good control over all 
  systematic errors \cite{davies03}.  One such systematic error is the 
  continued exclusion of heavy quark vacuum polarization.   State of the art lattice 
  simulations only include the effects of light ($s$,$u$ and $d$) 
  quark vacuum polarization.  

  The exclusion of heavy quark vacuum polarization is a good approximation because, 
  as we shall discuss, their effect on simulation results is suppressed 
  by a factor of $\bar{q}^{2}/m^{2}$, where $\bar{q}$ is a typical lattice 
  momentum.  For many lattice calculations (hadron spectra, decay constants) 
  $\bar{q} \approx \Lambda_{QCD} \approx 250 \textrm{MeV}$.  For a charm 
  quark $m_{c} \approx 1 \textrm{GeV}$ this could be as large as a 6\% effect.  
  Todays lattice calculations aspire to few percent ($1 - 3\%$) errors \cite{davies03}, 
  so for truly high-precision work, this error 
  should be removed.  It would be good to have a method for removing it 
  without having to do full dynamical simulations of the heavy quarks.  
  Fortunately these effects can be handled using effective field theory.

  \section{The Effective Lagrangian}

  The earliest example of an effective field theory is the Euler--Heisenberg 
  non-linear photon theory \cite{Heisenberg:1935qt}.  This is an expansion 
  of QED for low-energy photons with average energies $\omega \ll m_{e}$.  
  Euler and Heisenberg were concerned with non-linear interactions between 
  four photons, these corrections begin at $\mathcal{O}(\omega^{4}/m_{e}^{4})$,
  however there is a correction to the photon propagator which is a 
  $\mathcal{O}(\omega^{2}/m_{e}^{2})$ effect 
  \cite{Uehling:1935uj}.  Similar expansions apply to QCD.

  It is straightforward to write down the most general effective action
  that will account for these effects
  \begin{equation}
    \mathcal{S}_{EH} = \int d^{4}x \left[ 
      -\frac{1}{4} F_{a}^{\mu\nu}F^{a}_{\mu\nu} 
      + \sum_{Q} \left(\sum_{\mathcal{O}_{n}} 
      \frac{C_{\mathcal{O}_{n}}}{m_{Q}^{n+1}}\mathcal{O}_{n} \right)\right]
  \end{equation}
  where $\mathcal{O}_{n}$ are operators of dimension $2(n+1)$ built out of
  $F_{\mu\nu}$, $\tilde{F}_{\mu\nu}$ and $D_{\mu}$ that respect the symmetries 
  of QCD, $C_{\mathcal{O}_{n}}$ are the coefficients of these operators and the sum runs over all
  heavy quark flavors $Q=c,b,t$.  
  The operators of higher dimensions are suppressed by powers of 
  the heavy quark masses.  In this note, we will be concerned with only the 
  operators of dimension six.

  At dimension six there are two operators we need to consider \cite{simmons1}
  \begin{eqnarray}
    \mathcal{G}_{1} & = & g_{s}f_{abc}F^{\mu}_{a\nu}F^{\nu}_{b\sigma}F^{\sigma}_{c\mu} \label{op1} \\
    \mathcal{G}_{2} & = & \frac{1}{2}D_{\mu}F_{a}^{\mu\sigma}D_{\nu}F^{a}_{\nu\sigma} \label{op2}
  \end{eqnarray}
  With these operators the effective action is
  \begin{equation} \label{orig_act}
    S_{EH} = \int d^{4}x \left[ 
      -\frac{1}{4} F_{a}^{\mu\nu}F^{a}_{\mu\nu} 
      + \sum_{Q} \left(
      \frac{C_{\mathcal{G}_{1}}}{m_{Q}^{2}} \mathcal{G}_{1} +
      \frac{C_{\mathcal{G}_{2}}}{m_{Q}^{2}} \mathcal{G}_{2} \right) \right]
  \end{equation}
  This action can be matched to full QCD order by order in perturbation theory, 
  since a heavy quark loop is highly virtual.  Matching at leading 
  order gives the coefficients \cite{simmons2}
  \begin{equation}
    C_{\mathcal{G}_{1}} = -C_{F} \frac{\alpha_{s}}{360\pi}, \quad 
    C_{\mathcal{G}_{2}} = -C_{F} \frac{\alpha_{s}}{15\pi},
  \end{equation}
  where $C_{F} = 4/3$ for $SU(3)$.

  For matching to the lattice theory it is convenient to rewrite the action in terms
  of a different basis of operators.  We introduce the operator
  \begin{equation}
    \mathcal{G}_{3} = D^{\lambda}F_{a}^{\mu\nu} D_{\lambda}F^{a}_{\mu\nu},
  \end{equation}
  and make use of the relation \cite{simmons1}
  \begin{equation}
    \int d^{4}x \mathcal{G}_{1} = \int d^{4}x \left[ \frac{1}{2}\mathcal{G}_{3} - 2 \mathcal{G}_{2}\right]
  \end{equation}
  to rewrite the action as
  \begin{equation}
    S_{EH} = \int d^{4}x \left[ 
      -\frac{1}{4} F_{a}^{\mu\nu}F^{a}_{\mu\nu} 
      + \sum_{Q} \left(
      \frac{C_{\mathcal{G}_{1}}}{2m_{Q}^{2}} \mathcal{G}_{3} +
      \frac{
	\left(C_{\mathcal{G}_{2}} - 2 C_{\mathcal{G}_{1}}\right)
      }{
	m_{Q}^{2}
      } \mathcal{G}_{2} \right) \right].
  \end{equation}
  
  In this brief note we are primarily concerned with the effect of heavy quark 
  vacuum polarization on the gluon action.  With this in mind we make a field transformation
  \begin{equation}
    A^{a}_{\mu} \to A^{a}_{\mu} + 
    \frac{C}{m_{Q}^{2}} \alpha_{s}  D^{\nu} F^{a}_{\nu\mu}
  \end{equation}
  which gives
  \begin{equation}
    F^{a}_{\mu\nu}F_{a}^{\mu\nu} \to
    F^{a}_{\mu\nu}F_{a}^{\mu\nu}  -
    \frac{8C}{m_{Q}^{2}} \alpha_{s} \mathcal{G}_{2} + \mathcal{O}
    \left( \frac{1}{m_{Q}^{4}}\right).
  \end{equation}
  We can pick the constant $C$ to eliminate $\mathcal{G}_{2}$ from the gluon
  action\footnote{This technique is used to eliminate the same operator 
    in the lattice theory, with $m^{-2} \to a^{2}$.}.  This transformation will introduce
  additional four-quark operators which should be incorporated into the fermion actions used in 
  the simulations. 

  Finally, we rescale the fields $A \to A/g_{s}$, and express the action as
  \begin{eqnarray}\label{dim6eh_s}
    S_{EH} & = & -\frac{1}{2g_{s}^{2}} \int d^{4}x \left\lbrace
      \sum_{\mu\nu} \rm{Tr} \left[F_{\mu\nu}F_{\mu\nu} \right] \right. \nonumber \\
     &  + & \left. \sum_{Q} \frac{1}{m_{Q}^{2}}\left(
      C_{F} \frac{\alpha_{s}}{180\pi} \sum_{\mu\nu\lambda}
      \rm{Tr} \left[ D_{\lambda}F_{\mu\nu} D_{\lambda}F_{\mu\nu} \right]\right)\right\rbrace .
  \end{eqnarray}
  Our goal is to construct a lattice gluon action that reproduces this action.  Before
  we do this, it is useful to consider how large the mass $m_{Q}$ should be in order that
  the expansion in $m_{Q}^{-2}$ is sensible.

  \section{Limits on $m_{Q}$}

  In order to set limits on the expansion in $m_{Q}^{-2}$ we look at one part of the matching
  calculation for (\ref{orig_act}).  To match the coefficient $C_{\mathcal{G}_{2}}$ one
  can consider (light) quark-quark scattering at one loop order in continuum QCD.  The
  only effect of virtual heavy quarks at this order will be their contribution to the 
  gluon vacuum polarization.  Therefore, we can get a check on the validity of the
  expansion by investigating the one-loop gluon propagator in continuum QCD.

  In full QCD the renormalized gluon propagator is given (in Feynman gauge) by 
  \begin{equation}
    D_{\mu\nu}(q^{2}) = 
    \frac{-i g_{\mu\nu}}{q^{2}\left[1 - \hat{\Pi}(q^{2})\right]}.
  \end{equation}
  We separate the contribution of heavy quark loops out of the self energy
  \begin{equation}
    \hat{\Pi}(q^{2}) = \sum_{Q} \hat{\Pi}_{Q}(q^{2}) + \hat{\Pi}_{QCD}(q^{2}).
  \end{equation}
  The light quark and gluonic parts of the self energy will be described 
  by the standard QCD action, so we'll just retain the heavy quark 
  contribution.  Therefore, we will consider the propagator
  \begin{equation}
    D_{\mu\nu}(q^{2}) = \frac{-i g_{\mu\nu}}{q^{2}\left[
	1 - \sum_{Q} \hat{\Pi}_{Q}(q^{2})\right]}.
  \end{equation}
  The contribution to the self energy can be found in any textbook 
  (for example \cite{PeskinSchroeder}) it's just the QED contribution 
  times the color factor $C_{F}$.
  \begin{equation}\label{continuumprop1}
    \hat{\Pi}_{Q}(q^{2}) = - C_{F} \frac{2 \alpha_{s}}{\pi} 
    \int_{0}^{1} dx x(1-x) \log\left[
      \frac{1}{1 - \frac{q^{2}}{m_{Q}^{2}}x(1-x)}
      \right].
  \end{equation}
  Expanding in $q^{2}/m_{Q}^{2}$ we find
  \begin{equation} \label{continuumprop}
    \hat{\Pi}_{Q}(q^{2}) = -C_{F} \frac{\alpha_{s}}{15\pi} 
    \frac{q^{2}}{m_{Q}^{2}}.
  \end{equation}
  
  The range of validity for the expansion in $1/m_{Q}^{2}$ is set by the radius of convergence
  of the logarithim in (\ref{continuumprop1}),
  \begin{equation} \label{bound}
    |q| < 2 m_{Q}.
  \end{equation}
  Since our goal is to use this expansion in lattice simulations
  it is useful to express this in lattice units, multiplying both sides by 
  $a$, and taking the momentum $|q| = \Lambda_{QCD}$
  \begin{equation}
    m_{Q}a > 0.5 \Lambda_{QCD} a
  \end{equation}
  For a $0.1$ fm lattice $\Lambda_{QCD}a \approx 0.1 - 0.3$, so we get a very 
  loose limit, $m_{Q}a > 0.05 - 0.15$.  At this lattice spacing, 
  $m_{c}a = 0.5$, so this bound is satisfied.  Notice that for low 
  momentum processes it is not necessary for $m_{Q}a$ to be greater than 
  one, fundamentally this is an expansion in $|q|/m_{Q}$ \emph{not} 
  $1/(m_{Q}a)$.  For processes that involve larger internal momenta this 
  limit is more strict.  For example, the semileptonic form factor for
  $B \to \pi + \ell + \nu$ can involve internal momenta as high as
  \begin{equation}
    \sqrt{\frac{\Lambda_{QCD}m_{b}}{2}} \approx 1 \rm{GeV}.
  \end{equation}
  With $|q| = 1$ GeV (\ref{bound}) is still satisfied for a charm quark, but
  for lighter quarks it wouldn't be.  When we match to the lattice theory 
  we will find another bound on $m_{Q}a$.

  \section{Matching to the Lattice Theory}

  The leading correction in (\ref{dim6eh_s}) is trivially included in the widely used improved 
  gluon action 
  \begin{equation}
    S_{L} = \frac{2}{g_{0}^{2}} \sum_{x} \left\lbrace
    \beta_{P} \frac{1}{3} \textrm{Re}\rm{Tr}\left[1 - U_{P}\right] +
    \beta_{R} \frac{1}{3} \textrm{Re}\rm{Tr}\left[1 - U_{R}\right] +
    \beta_{6} \frac{1}{3} \textrm{Re}\rm{Tr}\left[1 - U_{6}\right] 
    \right\rbrace
  \end{equation}
  where $U_{P}$ is the plaquette, $U_{R}$ is the rectangle, and $U_{6}$ is the 
  parallelogram term\footnote{
    Corrections to the standard Wilson plaquette action due to heavy quark vacuum polarization 
    have been investigated in \cite{degrand,hasen}.}.  We define the operators
  \begin{eqnarray}
    \mathcal{F}_{0} & = & \sum_{\mu\nu} 
    \rm{Tr}\left(F_{\mu\nu}F_{\mu\nu}\right) \nonumber \\
    \mathcal{F}_{1} & = & \sum_{\mu\nu} 
    \rm{Tr}\left(D_{\mu}F_{\mu\nu}D_{\mu}F_{\mu\nu}\right) \\
    \mathcal{F}_{2} & = & \sum_{\mu\nu\sigma} 
    \rm{Tr}\left(D_{\sigma}F_{\mu\nu}D_{\sigma}F_{\mu\nu}\right) \nonumber
  \end{eqnarray}
  in terms of which we have
  \begin{eqnarray}
    \frac{1}{3} \textrm{Re}\rm{Tr}\left[1 - U_{P}\right]  & = & 
    -\frac{a^{4}}{4} \mathcal{F}_{0} + \frac{a^{6}}{24} 
    \mathcal{F}_{1} + \mathcal{O}\left(a^{8}\right) \nonumber \\
    \frac{1}{3} \textrm{Re}\rm{Tr}\left[1 - U_{R}\right]  & = & 
    -2 a^{4} \mathcal{F}_{0} + \frac{5 a^{6}}{6} \mathcal{F}_{1} + \mathcal{O}\left(a^{8}\right) \\
    \frac{1}{3} \textrm{Re}\rm{Tr}\left[1 - U_{6}\right]  & = & 
    -2 a^{4} \mathcal{F}_{0} + \frac{a^{6}}{3} \mathcal{F}_{1} 
    + \frac{a^{6}}{6} \mathcal{F}_{2} + \mathcal{O}\left(a^{8}\right). \nonumber
  \end{eqnarray}
  The choice\footnote{The factor of $u_{0}$ is a tadpole improvement 
    factor, it plays no role in our discussion.}
  \begin{eqnarray}
    \beta_{6} & = & 0 \nonumber \\
    \beta_{R} & = & -\frac{1}{20u^{2}_{0}} \beta_{P} \\
    \beta_{P} & = & \frac{5}{3} \nonumber 
  \end{eqnarray}
  gives \cite{Luscher:1984xn}
  \begin{equation}
    S_{L} = -\frac{1}{2g_{0}} \int d^{4}x \mathcal{F}_{0} + 
    \mathcal{O}\left(\alpha_{s}a^{2},a^{4},\alpha_{s}/(m_{c}a)^2\right).
  \end{equation}

  In order to reproduce (\ref{dim6eh_s}) we need to have 
  a non-zero value for $\beta_{6}$.  When we do this, we are be forced to 
  included an additional correction to $\beta_{R}$ in order to cancel any 
  potential contribution from $\mathcal{F}_{1}$.  We take
  \begin{eqnarray}
    \beta_{6} & = & \frac{\beta_{P}}{u_{0}^{2}} 
    \frac{\alpha_{s}}{(am_{Q})^{2}} C^{[1]} \nonumber \\
    \beta_{R} & = & -\frac{1}{20u^{2}_{0}} \beta_{P} 
    \left( 1 + \frac{\alpha_{s}}{(am_{Q})^{2}} R^{[1]}\right) \\
    \beta_{P} & = & \frac{5}{3} \nonumber
  \end{eqnarray}  
  which gives
  \begin{eqnarray} \label{latticeact}
    S_{L} & = &-\frac{1}{2g_{0}} a^{4} \sum_{x} \left\lbrace
    \left(1 - 
    \frac{\alpha_{s}}{(am_{Q})^{2}} \left[ \frac{40}{3}C^{[1]} - 
      \frac{2}{3}R^{[1]}\right]\right) \mathcal{F}_{0} 
    \right. \nonumber \\
    & + & a^{2} \frac{\alpha_{s}}{(am_{Q})^{2}} \left[\frac{5}{18}R^{[1]} 
      - \frac{10}{3}C^{[1]}\right] \mathcal{F}_{1} 
    \nonumber \\
    & - & \left. a^{2} \frac{10\alpha_{s}}{9(am_{Q})^{2}}C^{[1]} 
    \mathcal{F}_{2}\right\rbrace,
  \end{eqnarray}
  where for simplicity we just include only one flavor of heavy quark.  Setting
  \begin{equation}
    C^{[1]} = - C_{F} \frac{1}{200\pi} = - \frac{1}{150\pi}
  \end{equation}
  produces the correct coefficient for the $\mathcal{F}_{2}$ term.  
  The additional contribution to the coefficient
  of the $\mathcal{F}_{0}$ term produces an unobservable shift in the 
  wavefunction renormalization for the gluons, so
  we are free to drop it.  This leaves only the $\mathcal{F}_{1}$ term.  
  We tune $R^{[1]}$ to insure that
  its coefficient remains zero.  This is easily accomplished with 
  \begin{equation}
    R^{[1]} = 12 C^{[1]} = - \frac{2}{25\pi}.
  \end{equation}
  With these values the lattice action correctly reproduces 
  (\ref{dim6eh_s}) with 
  \begin{equation}
    \mathcal{O}\left(a^{4},\alpha_{s}a^{2},\alpha_{s}/(am_{c})^{4},\alpha_{s}^{2}/(am_{c})^{2}\right)
  \end{equation}
  errors.

  \section{Impact on One-Loop Corrections}
  
  We combine our results with the known one loop corrections to
  $\beta_{R}$ and $\beta_{6}$,
  \begin{eqnarray}
    \beta_{R} & = & -\frac{\beta_{P}}{20u_{0}^{2}} \left[
      1 + \alpha_{s}\left(\pi/a\right) \left( 0.4805 + X_{R} N_{f} - 
      \frac{2}{25\pi} \sum_{Q} \frac{1}{(am_{Q})^{2}} \right) 
      \right] \label{finalbetar} \\
    \beta_{6} & = & -\frac{\beta_{P}}{u_{0}^{2}} \alpha_{s}\left(\pi/a\right) \left( 
    0.03325 + X_{6} N_{f} + 
    \frac{1}{150\pi} \sum_{Q} \frac{1}{(am_{Q})^{2}} \right).
  \end{eqnarray}
  The one loop terms due to the gluons are well 
  known \cite{Luscher:1985zq,alford95}, those due to dynamical light fermions
  are being computed \cite{mason05}.  With $am_{c} \approx 1$ these 
  corrections to the one-loop terms are 
  $0.025$ for $\beta_{R}$ and $0.0021$ for $\beta_{6}$.  
  
  These expressions also limit how small one can take $a$.  The coupling
  constant $\alpha_{s}\left(\pi/a\right)$ goes to zero logarithmically as $a\to 0$, 
  whereas the correction from the heavy quark vacuum polarization is growing as $1/a^{2}$. 
  Clearly, if we take $a$ too small the correction term will end up 
  larger than the tree level term.  For example, on an ultra-fine lattice,
  with $a=0.01$ fm, we have $m_{c}a = 0.05$, which gives (setting $N_{f}=0$)
  \begin{equation}
    \beta_{R} = -\frac{\beta_{P}}{20 u_{0}^{2}} \left[1 - 9.7\alpha_{s}\left(\pi/a\right)\right], 
    \quad
    \beta_{6} = -\frac{\beta_{P}}{u_{0}^{2}} 0.88\alpha_{s}\left(\pi/a\right).
  \end{equation}
  In this case, the ``correction'' to $\beta_{R}$ is larger than the leading order 
  term.  With $m_{c}a = 0.13$ the coefficient of the correction term becomes one, so we 
  take this as a rough lower limit, $a=0.025$ fm.  

  A more realistic lattice spacing is $a = 0.09$ fm, which gives 
  $m_{c}a = 0.45$ and
  \begin{eqnarray}
    \beta_{R} & = & -\frac{\beta_{P}}{20 u_{0}^{2}} \left[1 + (0.481 - 0.126)\alpha_{s}\left(\pi/a\right)\right] = 
    -\frac{\beta_{P}}{20 u_{0}^{2}} \left[1 + 0.355\alpha_{s}\left(\pi/a\right)\right] , \nonumber \\
    \beta_{6} & = & -\frac{\beta_{P}}{u_{0}^{2}} (0.0333 + 0.0105)\alpha_{s}\left(\pi/a\right) = 
    -\frac{\beta_{P}}{u_{0}^{2}} 0.0437\alpha_{s}\left(\pi/a\right).
  \end{eqnarray}
  In the this case, the corrections due to heavy quark
  vacuum polarization are significant but do not cause convergence problems.

  \section{Conclusion}
  
  In this note we have demonstrated that heavy quark vacuum polarization can be 
  included in Monte Carlo lattice simulations with no additional computational 
  cost.  The action is the standard Symanzik improved gluon action
  \begin{equation}
    S_{L} = \frac{2}{g_{0}^{2}} \sum_{x} \left\lbrace
    \beta_{P} \frac{1}{3} \textrm{Re}\rm{Tr}\left[1 - U_{P}\right] +
    \beta_{R} \frac{1}{3} \textrm{Re}\rm{Tr}\left[1 - U_{R}\right] +
    \beta_{6} \frac{1}{3} \textrm{Re}\rm{Tr}\left[1 - U_{6}\right] 
    \right\rbrace
  \end{equation}
  with modified coefficients
  \begin{eqnarray}
    \beta_{R} & = & -\frac{\beta_{P}}{20u_{0}^{2}} \left[
      1 + \alpha_{s}\left(\pi/a\right) \left( 0.4805 + X_{R} N_{f} - 
      \frac{2}{25\pi} \sum_{Q} \frac{1}{(am_{Q})^{2}} \right) \right] 
    \nonumber \\
    \beta_{6} & = & -\frac{\beta_{P}}{u_{0}^{2}} \alpha_{s}\left(\pi/a\right) \left( 
    0.03325 + X_{6} N_{f} + 
    \frac{1}{150\pi} \sum_{Q} \frac{1}{(am_{Q})^{2}} \right)  \\
    \beta_{P} & = & \frac{5}{3}. \nonumber
  \end{eqnarray}
  This action reproduces continuum QCD up to corrections of order 
  $a^{4}$,$(\alpha_{s}a)^{2}$, $\alpha_{s}/(am_{c})^{4}$ and 
  $\alpha_{s}^{2}/(am_{c})^{2}$.  In order to insure reasonable behaviour
  of the perturbation series $m_{c}a \gtrsim 0.13$ is required.  With this 
  constraint, this action should simulate processes with average internal
  momentum $|q| \ll 2 m_{c}$.  Ideally, the charm quark would be
  treated dynamically, then these constraints would apply $m_{b}a$ 
  instead.
  
  \section*{Acknowledgments}
  
  We thank Peter Lepage for many useful discussions and Quentin Mason for a discussion
  regarding the improved action. This
  work was supported in part by the National Science Foundation under
  grant number PHY0098631.

\end{document}